\documentclass{elsart}
\usepackage{epsfig}
\usepackage{amssymb}
\usepackage{amsmath}

\usepackage{color}

\newcommand{\affuni}[2]{Dipartimento di Fisica dell'Universit\`a #1, #2, Italy.}
\newcommand{\affinfn}[2]{INFN Sezione di #1, #2, Italy.}

\newcommand{\dd}{\mathrm{d}}

\begin{document}

\begin{frontmatter}

\title{\boldmath Measurement of the $\phi \to \pi^0  e^+e^-$ transition form factor with the KLOE detector}

\collab{The KLOE-2 Collaboration}
\author[Messina,Frascati]{A.~Anastasi},
\author[Frascati]{D.~Babusci\corauthref{cor}},
\ead{danilo.babusci@lnf.infn.it}
\author[Frascati]{G.~Bencivenni},
\author[Warsaw]{M.~Berlowski},
\author[Frascati]{C.~Bloise},
\author[Frascati]{F.~Bossi},
\author[INFNRoma3]{P.~Branchini},
\author[Roma3,INFNRoma3]{A.~Budano},
\author[Uppsala]{L.~Caldeira~Balkest\aa hl},
\author[Uppsala]{B.~Cao},
\author[Roma3,INFNRoma3]{F.~Ceradini},
\author[Frascati]{P.~Ciambrone},
\author[Messina,INFNCatania,Novosibirsk]{F.~Curciarello},
\author[Cracow]{E.~Czerwi\'nski},
\author[Roma1,INFNRoma1]{G.~D'Agostini},
\author[Frascati]{E.~Dan\`e},
\author[INFNRoma3]{V.~De~Leo},
\author[Frascati]{E.~De~Lucia},
\author[Frascati]{A.~De~Santis},
\author[Frascati]{P.~De~Simone},
\author[Roma3,INFNRoma3]{A.~Di~Cicco},
\author[Roma1,INFNRoma1]{A.~Di~Domenico},
\author[INFNRoma2]{R.~Di~Salvo},
\author[Frascati]{D.~Domenici},
\author[Frascati]{A.~D'Uffizi},
\author[Roma2,INFNRoma2]{A.~Fantini},
\author[Frascati]{G.~Felici},
\author[ENEACasaccia,INFNRoma1]{S.~Fiore},
\author[Cracow]{A.~Gajos},
\author[Roma1,INFNRoma1]{P.~Gauzzi},
\author[Messina,INFNCatania]{G.~Giardina},
\author[Frascati]{S.~Giovannella},
\author[INFNRoma3]{E.~Graziani},
\author[Frascati]{F.~Happacher},
\author[Uppsala]{L.~Heijkenskj\"old},
\author[Uppsala]{W.~Ikegami Andersson},
\author[Uppsala]{T.~Johansson},
\author[Cracow]{D.~Kami\'nska},
\author[Warsaw]{W.~Krzemien},
\author[Uppsala]{A.~Kupsc},
\author[Roma3,INFNRoma3]{S.~Loffredo},
\author[Messina2,INFNMessina] {G.~Mandaglio},
\author[Frascati,Marconi]{M.~Martini},
\author[Frascati]{M.~Mascolo\corauthref{cor}},
\ead{mascolo.matteo@gmail.com}
\author[Roma2,INFNRoma2]{R.~Messi},
\author[Frascati]{S.~Miscetti},
\author[Frascati]{G.~Morello},
\author[INFNRoma2]{D.~Moricciani},
\author[Cracow]{P.~Moskal},
\author[Uppsala]{M.~Papenbrock},
\author[INFNRoma3]{A.~Passeri},
\author[Energetica,INFNRoma1]{V.~Patera},
\author[Frascati]{E.~Perez~del~Rio},
\author[INFNBari]{A.~Ranieri},
\author[Cracow]{P.~Salabura}
\author[Frascati]{P.~Santangelo},
\author[Frascati]{I.~Sarra},
\author[Calabria,INFNCalabria]{M.~Schioppa},
\author[Frascati]{M.~Silarski},
\author[Frascati]{F.~Sirghi},
\author[INFNRoma3]{L.~Tortora},
\author[Frascati]{G.~Venanzoni},
\author[Warsaw]{W.~Wi\'slicki},
\author[Uppsala]{M.~Wolke}
\address[INFNBari]{\affinfn{Bari}{Bari}}
\address[INFNCatania]{\affinfn{Catania}{Catania}}
\address[Cracow]{Institute of Physics, Jagiellonian University, Cracow, Poland.}
\address[Frascati]{Laboratori Nazionali di Frascati dell'INFN, Frascati, Italy.}
\address[Messina]{Dipartimento di Fisica e Scienze della Terra dell'Universit\`a di Messina, Messina, Italy.}
\address[Messina2]{Dipartimento di Scienze Chimiche, Biologiche, Farmaceutiche ed Ambientali dell'Universit\`a 
di Messina, Messina, Italy.}
\address[INFNMessina]{INFN Gruppo collegato di Messina, Messina, Italy.}
\address[Calabria]{\affuni{della Calabria}{Rende}}
\address[INFNCalabria]{INFN Gruppo collegato di Cosenza, Rende, Italy.}
\address[Energetica]{Dipartimento di Scienze di Base ed Applicate per l'Ingegneria dell'Universit\`a 
``Sapienza'', Roma, Italy.}
\address[Marconi]{Dipartimento di Scienze e Tecnologie applicate, Universit\`a ``Guglielmo Marconi", Roma, Italy.}
\address[Novosibirsk]{Novosibirsk State University, 630090 Novosibirsk, Russia.}
\address[Roma1]{\affuni{``Sapienza''}{Roma}}
\address[INFNRoma1]{\affinfn{Roma}{Roma}}
\address[Roma2]{\affuni{``Tor Vergata''}{Roma}}
\address[INFNRoma2]{\affinfn{Roma Tor Vergata}{Roma}}
\address[Roma3]{Dipartimento di Matematica e Fisica dell'Universit\`a ``Roma Tre'', Roma, Italy.}
\address[INFNRoma3]{\affinfn{Roma Tre}{Roma}}
\address[ENEACasaccia]{ENEA UTTMAT-IRR, Casaccia R.C., Roma, Italy}
\address[Uppsala]{Department of Physics and Astronomy, Uppsala University, Uppsala, Sweden.}
\address[Warsaw]{National Centre for Nuclear Research, Warsaw, Poland.}
\corauth[cor]{Corresponding authors}

\begin{abstract}
A measurement of the vector to pseudoscalar conversion decay $\phi \to \pi^0 e^+e^-$ with the KLOE experiment is presented. A sample of $\sim 9500$ 
signal events was selected from a data set of 1.7 fb$^{-1}$ of $e^+e^-$ collisions at $\sqrt{s} \sim m_{\phi}$ collected at the DA$\Phi$NE $e^+e^-$ collider. 
These events were used to perform the first measurement of the transition form factor $| F_{\phi \pi^0}(q^2) |$ and a new measurement of the branching 
ratio of the decay: $\rm{BR}\,(\phi \to \pi^0 e^+e^-) = (\,1.35 \pm  0.05^{\,\,+0.05}_{\,\,-0.10}\,) \times 10 ^{-5}$. The result improves significantly on previous 
measurements and is in agreement with theoretical predictions. 
\end{abstract}


\begin{keyword}
$e^+e^-$ Collisions \sep Conversion Decay \sep Transition Form Factor
\PACS 13.66.Bc, 13.40.Gp 
\end{keyword}

\end{frontmatter}


\section{Introduction}\label{Sec:Intro}
The conversion decays of a light vector resonance (V) into a pseudoscalar meson (P) and a lepton pair, $V \to P \,\gamma^* \to P\,\ell^+ \ell^-$, represent a 
stringent test for theoretical models of the nature of mesons. In these processes, the squared dilepton invariant mass, $m_{\ell\ell}^2$, corresponds to the 
virtual photon 4-momentum transfer squared, $q^2$. The $q^2$ distribution depends on the underlying electromagnetic dynamical structure of the transition 
$V \to P \,\gamma^*$.  

The description of the coupling of the mesons to virtual photons is typically parametrized by the so-called Transition Form Factor (TFF), $F_{VP}(q^2)$. 
TFFs are fundamental quantities playing an important role in many fields of particle physics, such as the calculation of the hadronic Light-by-Light contribution 
to the Standard Model prediction of the muon anomalous magnetic moment \cite{bib:gminustwo}.

Recently, the increasing interest in conversion decays was mostly driven by the discrepancy between the experimental data from NA60 \cite{bib:na60} and 
Lepton G \cite{bib:lepg}, and the Vector Meson Dominance (VMD) prediction for the $\omega \to \pi^0 \mu^+ \mu^-$ TFF $F_{\omega\,\pi^0} (q^2)$. Over 
the years, several theoretical models have been developed to explain this discrepancy \cite{bib:Leupold, bib:SKN,bib:iva, bib:dan}. In order to check the 
consistency of the models, a measurement of the $F_{\phi \pi^0}(q^2) $ TFF, which has never been measured so far, was strongly recommended. In particular, 
because of its kinematics, the $\phi \to \pi^0 e^+e^-$ process is a very good benchmark to investigate the observed steep rise in NA60 data at $q^2$ close to 
the $\rho$ resonance mass.

At present, the existing data on $\phi \to \pi^0 e^+e^-$ come from SND \cite{bib:SND} and CMD-2 \cite{bib:CMD2} experiments which were able to extract only 
the value of the Branching Ratio (BR). The $F_{\phi \, \pi^0} (q^2)$ TFF hence, was never measured so far. Its modulus square enters in the calculation of the 
$\phi \to \pi^0 e^+e^-$ double-differential decay width:
\begin{equation}
\frac{\dd^2 \Gamma(\phi \to \pi^0 e^+e^-)}{\dd q^2\,\dd \cos \theta^*} = \frac38\,\left ( \frac{q^2}{q^2 + 2m_e^2}\right)\,
(2 - \beta^2\,\sin^2 \theta^*)\,\frac{\dd \Gamma(\phi \to \pi^0 e^+e^-)}{\dd q^2} 
\label{eq:decw}
\end{equation} 
with $\beta = \left ( 1 - 4m_e^2/q^2 \right)^{1/2}$ and \cite{bib:lands}:
\begin{align} 
\frac{\dd \Gamma(\phi \to \pi^0 e^+e^-)}{\dd q^2}  &= \Gamma (\phi \to \pi^0 \gamma)\,\frac{\alpha}{3\,\pi}\,\beta\,\frac{| F_{\phi \pi^0}(q^2) |^2}{q^2}\,
\left (1 + \frac{2 m_e^2}{q^2}\right) \times \nonumber \\[8pt] 
& \qquad \left[\left(1 + \frac{q^2}{m_\phi^2 - m_\pi^2}\right)^2 - \frac{4 m_\phi^2 q^2}{(m_\phi^2 - m_\pi^2)^2}\right ]^{3/2},
\label{eq:lands}
\end{align}
where $m_e$ is the mass of the electron, and $m_\phi$, $m_\pi$ are the masses of the $\phi$ and $\pi^0$ mesons, respectively. $\theta^*$ is the angle between 
the $\phi$ and the $e^+$ direction in the $e^+e^-$ rest frame. Its cosine is an invariant quantity which can be written as \cite{bib:iva_priv}:
\begin{equation}
\cos \theta^* = \frac{(q^2 + m_{\phi}^2 - m_{\pi}^2) - 4\,p_{\phi} \cdot p_{e^+}}{\beta \sqrt{\left (q^2 - m_{\phi}^2 - m_{\pi}^2 \right )^2 - 4\,m_{\pi}^2\,m_{\phi}^2}},
\label{eq:cos}
\end{equation}
where $p_{\phi}$  is the 4-momentum of $\phi$ and $p_{e^+}$ of the positron. 

Thanks to the large amount of collected $\phi$ decays ($\sim 5.6 \times 10^9$), the KLOE experiment has been able both to perform the first measurement 
of the $F_{\phi \, \pi^0} (q^2)$ TFF and to significantly improve the determination of the branching ratio of $\phi \to \pi^0 e^+e^-$.

\section{The KLOE detector}\label{sec:kloe}
DA$\Phi$NE, the Frascati $\phi$-factory, is an $e^+e^-$ collider running at a center-of-mass energy of $\sim 1020$~MeV. Positron and electron beams 
collide at an angle of $\pi$-25 mrad, producing $\phi$ mesons nearly at rest.  

The KLOE apparatus consists of a large cylindrical Drift Chamber (DC) surrounded by a lead-scintillating fiber electromagnetic calorimeter both inserted 
inside a superconducting coil, providing a 0.52 T axial field. The beam pipe at the interaction region is a sphere with 10 cm radius, made of a 0.5 mm thick 
Beryllium-Aluminum alloy. The drift chamber \cite{bib:DC}, 4 m in diameter and 3.3 m long, has 12,582 all-stereo tungsten sense wires and 37,746 aluminum 
field wires, with a shell made of carbon fiber-epoxy composite with an internal wall of $\sim 1$ mm thickness. The gas used is a 90\% helium, 10\% isobutane 
mixture. The momentum resolution is $\sigma(p_{\perp})/p_{\perp}\approx 0.4\%$. Vertices are reconstructed with a spatial resolution of $\sim$ 3~mm. The 
calorimeter \cite{bib:EMC}, with a readout granularity of $\sim$\,(4.4 $\times$ 4.4) cm$^2$, for a total of 2440 cells arranged in five layers, covers 98\% of the 
solid angle. Each cell is read out at both ends by photomultipliers, both in amplitude and time. The energy deposits are obtained from the signal amplitude 
while the arrival times and the particle positions are obtained from the time of the signals collected at the two ends. Cells close in time and space are 
grouped into energy clusters. Energy and time resolutions are 
$\sigma_E/E = 5.7\%/\sqrt{E\ {\rm(GeV)}}$ and  $\sigma_t = 57\ {\rm ps}/\sqrt{E\ {\rm(GeV)}} \oplus100\ {\rm ps}$, respectively. The trigger \cite{bib:TRG} 
uses both calorimeter and chamber information. In this analysis the events are selected by the calorimeter trigger, requiring two energy deposits with 
$E > 50$ MeV for the barrel and $E > 150$ MeV for the endcaps.

Large angle Bhabha scattering events are used to obtain luminosity, center-of-mass energy and crossing angle of the beams. A precision measurement of $\sqrt{s}$, 
with negligible statistical uncertainty and a systematic error of $\sim$ 30 keV, is routinely performed on the basis of  200 nb$^{-1}$ of integrated luminosity. 
The systematic error is in fact on the absolute momentum scale, derived from the analysis of the $\phi$ lineshape \cite{bib:mks}. The center-of-mass energy distribution 
width is about 330 keV from the contributions of i) DA$\Phi$NE beam energy spread (0.06\%) and ii) radiative corrections/effects. Collected data are processed by an 
event classification algorithm \cite{bib:offline}, which streams various categories of events in different output files.

\section{Data analysis}
The analysis of the decay $\phi \to \pi^0 e^+e^-$ ($\pi^0 \to \gamma\gamma$), has been performed on a data sample of $1.69$ fb$^{-1}$ from the 2004/2005 
data taking campaign. 

The simulation of both signal and background events is based on the KLOE Monte Carlo (MC), {\tt GEANFI} \cite{bib:offline}, that includes radiative contributions 
to the process under study and takes into account variations of beam energy, crossing angle and machine background conditions on a run-by-run basis. 
The MC simulation of the signal has been produced according to Eq. \eqref{eq:decw}, assuming a point-like TFF (i.e. $|F_{\phi \, \pi^0} (q^2)|^2 = 1$). The 
radiative emission from the leptons in the final state of the channel under study is also included in the simulation by means of the {\tt PHOTOS} MC generator 
\cite{bib:photos}. The signal production corresponds to an integrated luminosity 1000 times larger than for the collected data. The dominant contributions to 
background events originate from double radiative Bhabha scattering ($e^+e^- \to e^+e^- \gamma \gamma$) and from the $\phi \to \pi^0 \gamma$ decay, 
where the $\gamma$ converts to a $e^+e^-$ pair in the interaction with the beam pipe or drift chamber walls. (The $\phi \to \pi^0 \gamma$ with the $\pi^0$ Dalitz 
decay to $\gamma e^+e^-$ also contributes to the background but it is almost completely suppressed by the  analysis cuts). All other background events, i.e. the other 
$\phi$ meson decays, the non-resonant $e^+e^- \to \omega\pi^0$ process and the $\pi^0$ production via $\gamma\gamma$ interaction,  $e^+e^- \to \pi^0 e^+e^-$, were 
also simulated, resulting fully negligible at the end of the analysis path. 

As a first step of the analysis, events are selected requiring two opposite-charge tracks extrapolated to a cylinder around the interaction point (IP) with radius 
4 cm and 20 cm long and two prompt photon candidates from IP (i.e. with energy clusters $E_{\rm clu} > 7$ MeV not associated to any track, in the angular 
region $| \cos \theta_\gamma | < 0.92$ and in the time window $| T_\gamma - R_\gamma/c | < \mathrm{min}\,(3 \sigma_t, 2\,{\rm ns}))$. In order to enhance 
the signal-to-background ratio, further constraints are applied on this preselected data sample:  
\begin{itemize}
\item a cut on the energies of the final state particles requiring: ($30 <  E_{e^{\pm}} < 460$)~MeV, $E_{\gamma} > 70$~MeV, 
($300 < E_{\gamma_1} + E_{\gamma_2} <  670$)~MeV and ($470 <  E_{e^+} + E_{e^-} < 750$)~MeV;
\item angular cuts: $45^\circ < \theta_{e^\pm}, \theta_\gamma < 135^\circ$, $\theta_{e^+e^-} < 145^\circ$ and 
$27^\circ <  \theta_{\gamma\gamma} < 57^\circ$;
\item two cuts on the invariant mass of the two photons and on the recoil mass against $e^+e^-$ to select events with a $\pi^0$ in the 
final state, i.e. ($90 < m_{\gamma\gamma}^{\,\rm inv} < 190$)~MeV and ($80 < m_{e^+e^-}^{\,\rm miss} < 180$)~MeV;
\item a cut on the invariant mass and the distance between the two tracks calculated at the surfaces of the beam pipe (BP) or of the DC wall surfaces;
\item a cut based on the time of flight (ToF) of the tracks to the calorimeter.
\end{itemize}

All the cuts have been optimized in order to maximize the available range of the $e^+ e^-$ invariant mass spectrum for the TFF extraction. The constraints on 
angular and energy variables have been obtained looking at the differences between the signal and Bhabha reconstructed angular and energy distributions 
of final leptons and photons. The cuts on the energies and on the opening angles $\theta_{e^+e^-}$ and $\theta_{\gamma\gamma}$ of tracks and clusters allow 
to strongly suppress the dominant background (S/B $\sim 5 \times 10^{-4}$) from the QED process $e^+e^- \to e^+e^- \gamma \gamma$. 
The $\theta_{\, e^+e^-} \leq  145^\circ$ requirement is also very effective in rejecting of the irreducible background from the $\gamma\gamma$ process 
$e^+e^- \to e^+e^- \pi^0$, in which the final state leptons are emitted in the forward direction (i.e. at small polar angles with respect to the beam line) for this 
kind of events. The $\phi \to \pi^0\gamma$ contamination, with the $\gamma$ converting on the BP or DC walls, is suppressed by tracing back the tracks of the 
$e^+/e^-$ candidates, by reconstructing the invariant mass ($m^{\,\rm{BP,DC}}_{e^+e^-}$) and the distance ($d^{\,\rm{BP,DC}}_{e^+e^-}$) of the track pair both 
at the BP and DC wall surfaces. Both variables are expected to be small for photon conversion events, so that this background is suppressed by rejecting events 
with: $m_{e^+e^-}^{\,\rm{BP}} < 10$ MeV and $d_{e^+e^-}^{\,\rm{BP}} < 2$ cm, or $m_{e^+e^-}^{\,\rm{DC}} < 80$~MeV and $d_{e^+e^-}^{\,\rm{DC}} < 3$ cm. 
The cut on the time of flight to the calorimeter is used to remove residual background events with muons or charged pions in the final state. When an energy 
cluster is associated to a track, the ToF to the calorimeter is evaluated using both the calorimeter timing ($t_{\rm{clu}}$) and the time along the track trajectory, 
namely $t_{\rm{trk}} = L_{\rm{trk}}/\beta c$, where $L_{\rm{trk}}$ is the length of the track path. The difference $\Delta t = t_{\rm{trk}} - t_{\rm{clu}}$ is then 
evaluated in the electron hypothesis; all events with $\Delta t  < 0.8$ ns are retained for further analysis. This algorithm, together with the cut on the energies 
of the final particles, turns out to be crucial for reducing the contamination from the decay $\phi \to \pi^+ \pi^- \pi^0$ to a negligible level.

After all the above described cuts the overall efficiency, as estimated by the MC, is 15.4\%. The efficiency is 19.5\% at lower $e^+e^-$ invariant masses, 
decreasing to a few percent at the highest values of momentum transfer. For this reason the analysis is limited up to $\sqrt{q^2} = 700$ MeV. At the end of the 
analysis chain, 14670 events are selected, with a residual background contamination of $\sim 35 \%$, equally divided between the Bhabha and 
$\phi \to \pi^0\gamma$ component, corresponding to about 9500 signal events. 

\begin{figure}[!ht]
  \begin{center}
   \begin{tabular}{cc}
      \hspace{-0.8 cm}
      \epsfig{file=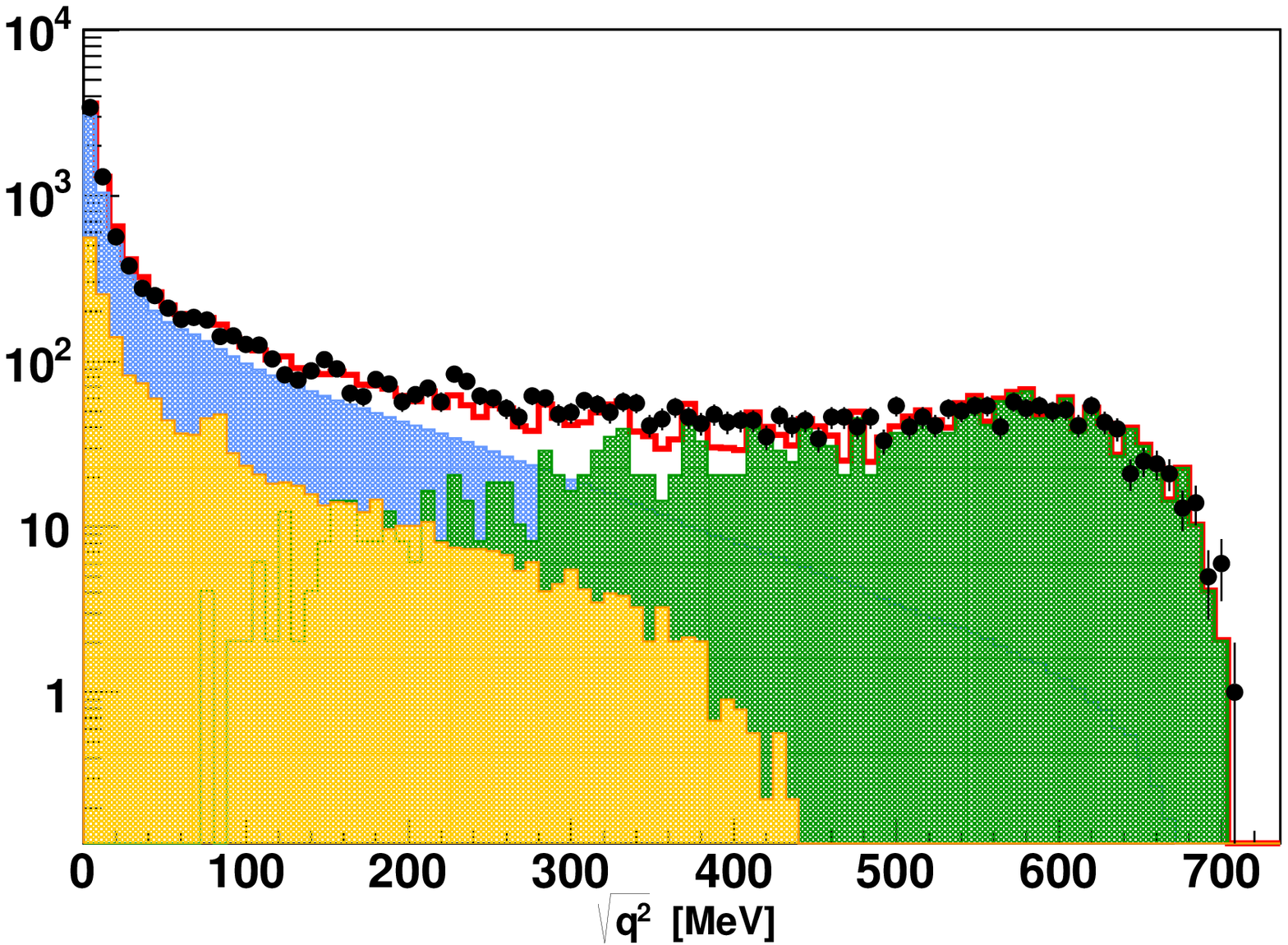,width=0.52\textwidth}    
      \epsfig{file=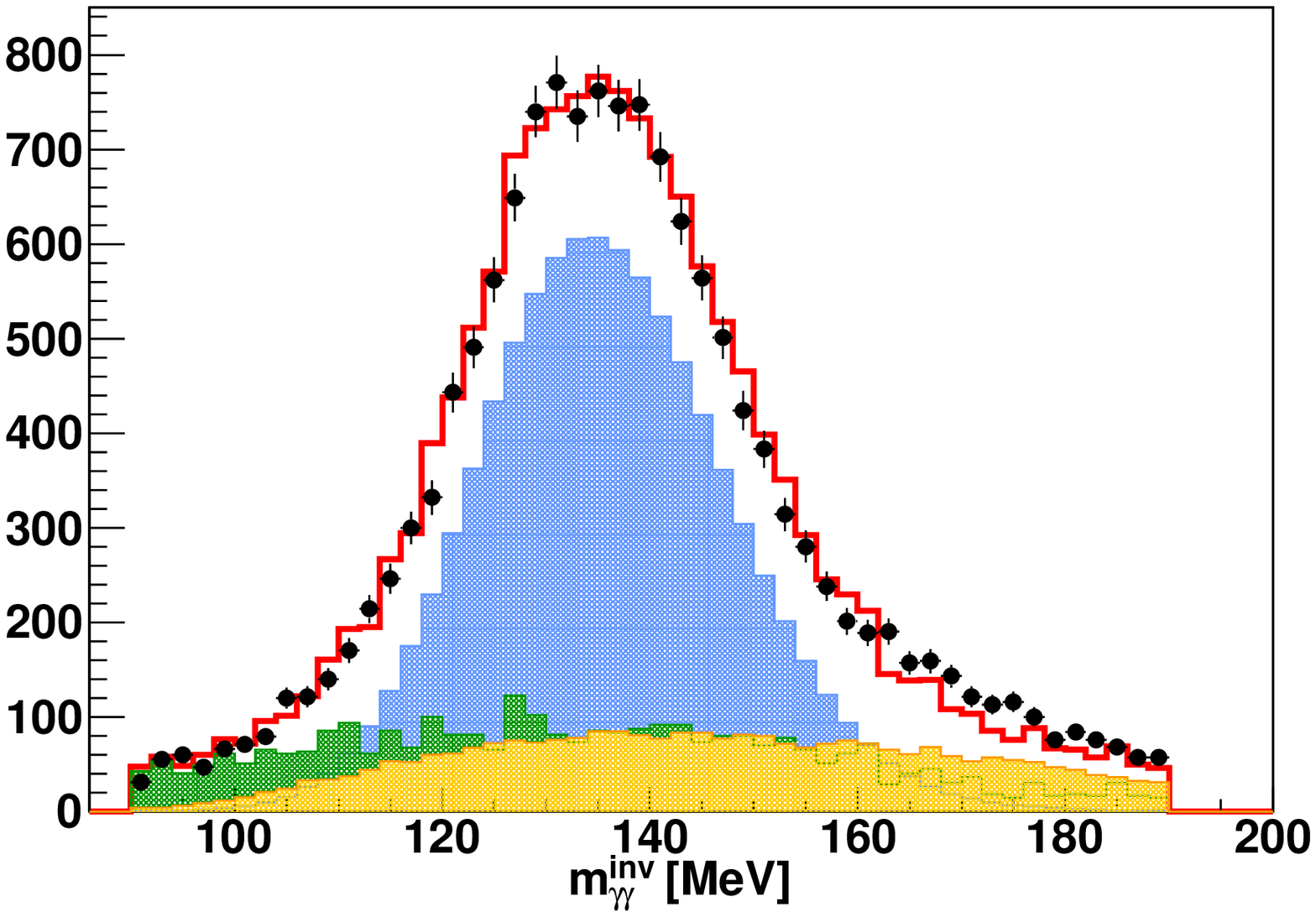,width=0.52\textwidth}
      \hspace{-0.5 cm}
   \end{tabular}
  \end{center}
  \caption{Data-MC comparison after all the analysis cuts for the invariant-mass spectrum of $e^+e^-$ (left)
  and of the two photons (right). Black dots are data, solid red line is the sum of MC histogram components: 
  signal (cyan), $\phi \to \pi^0 \gamma$ background (orange) and radiative Bhabha scattering (green).}
  \label{fig:agree}
\end{figure}

\begin{figure}[!ht]
  \begin{center}
   \begin{tabular}{cc}
      \epsfig{file=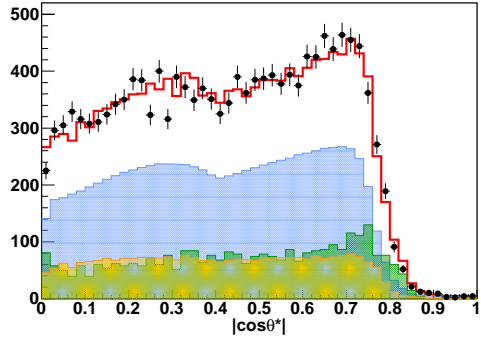,width=0.52\textwidth}
   \end{tabular}
  \end{center}
  \caption{Data-MC comparison after all the analysis cuts for $| \cos \theta^* |$. Code of symbols and colors as in Fig. \ref{fig:agree}.}
  \label{fig:costh}
\end{figure}

The agreement between data and Monte Carlo simulation, after all selection cuts, is shown in Fig. \ref{fig:agree} for the $\sqrt{q^2}$ and $m_{\gamma\gamma}$ 
distributions. As shown in the left panel of this Figure, in the region $\sqrt{q^2} > 400$ MeV the $\phi \to \pi^0\gamma$ background is negligible and only 
the Bhabha background is present. Furthermore, as a check of Eq. \eqref{eq:cos}, in Fig. \ref{fig:costh} we show the distribution of $| \cos \theta^* |$ as compared 
to the MC prediction.

\begin{figure}[!ht]
  \begin{center}
      \epsfig{file=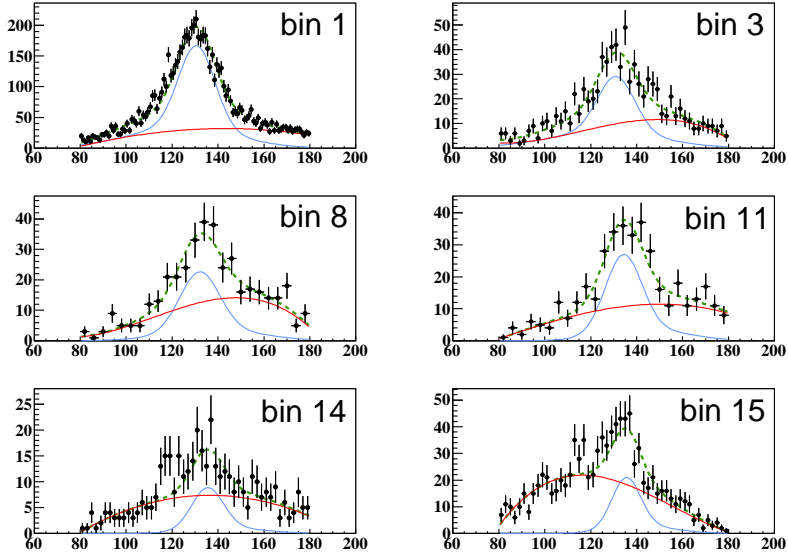, width=0.8\textwidth}    
  \end{center}
  \caption{$m^{\rm{miss}}_{e^+e^-}$ distributions (units MeV) for some $\sqrt{q^2}$ bins showing the total 
  background contribution (red curve) evaluated from a fit to the data (black points), with fixed signal shape 
  (blue curve). The dashed green curve represents the global fit of data, including the background function 
  and the signal parametrization.}
  \label{fig:bckg}
\end{figure}

In order to subtract the residual background from data, the $e^+e^-$ invariant-mass spectrum is divided into 15 bins of increasing width (to preserve the 
statistics of signal candidates). In each bin of $\sqrt{q^2}$, the $m_{e^+e^-}^{\,\rm{miss}}$ distribution is fit by a sum of two Gaussian functions, parametrizing 
the signal, and a third-order polynomial, parametrizing the background. Some examples of the fits to the $m_{e^+e^-}^{\,\rm{miss}}$ distributions are shown in 
Fig. \ref{fig:bckg}.  Apart from a global normalization, the parameters of the Gaussian functions are fixed by a fit of the MC signal distribution. The background 
contribution is evaluated bin by bin, without any assumption or constraint for the polynomial parameters. Once the residual background is parametrized, it is 
bin by bin subtracted from data. 

\subsection{Measurement of $| F_{\phi \, \pi^0} (q^2) |^2$}\label{s:tff}
The modulus square of the TFF, $| F_{\phi\pi^0}(q^2) |^2$, is a factor in front of the $q^2$ differential cross section (see Eq. \eqref{eq:lands}), hence it can be 
extracted from data by dividing the measured $e^+e^-$ invariant-mass spectrum by the spectrum of reconstructed MC signal events, generated with a constant 
$F_{\phi \, \pi^0}(q^2)$, after all the analysis cuts. The result is reported in Table \ref{tab:TFF}. The measured TFF is normalized so that 
$| F_{\phi \, \pi^0}(q^2) |^2 = 1$ in the first bin. The errors include both the statistical and the systematic uncertainty. 
\begin{table}[!ht]
\centering
\caption{KLOE measurement of the transition form factor $| F_{\phi \, \pi^0}(q^2) |$ of the $\phi \to \pi^0 e^+e^-$ decay.}
\bigskip
\begin{tabular}{ c c c c c }
\hline  
Bin \# \,\quad & $\sqrt{q^2}$-range\,\quad & Bin \!center\,\quad & $\sqrt{q^2}$ (UChT) \,\quad & $| F_{\phi \, \pi^0} (q^2) |^2$  \\  
   & \quad (MeV) \quad &  \quad (MeV) \quad &  \quad (MeV) \quad & \\
   \hline
1 & $2m_e \div 30$   & 15.5 & 9.0 &  \, 1.00 $\pm$  0.11 \, \\
2 & $30 \div 60$   & 45  & 43.3 & \, 1.18 $\pm$  0.22 \, \\
3 & $60 \div 90$        & 75   & 74.0 & \, 0.93 $\pm$ 0.21 \, \\
4  & $90 \div 120$     &105 & 104.2 & \, 1.09 $\pm$ 0.19 \, \\
5  &  $120 \div 150$  & 135  & 134.4 & \, 1.19 $\pm$ 0.23 \, \\
6  &  $150 \div 190$    & 170   &  169.0 &\, 1.42 $\pm$ 0.33 \, \\
7  &  $190 \div 230$    &  210  & 209.1 &\, 1.46 $\pm$  0.47 \, \\
8  &  $230 \div 270$    & 250  & 249.1 &\, 1.22 $\pm$ 0.58 \, \\
9  &  $270 \div 310$     & 290 & 288.8 &\, 2.30 $\pm$ 0.53 \, \\
10  &  $310 \div 350$   & 330 & 327.5 &\, 2.17 $\pm$  0.65 \, \\
11  &  $350 \div 400$    &  375 & 380.0 &\, 3.01 $\pm$ 1.34 \, \\
12  &  $400 \div 450$    & 425 & 426.6 &\, 3.14 $\pm$  1.71 \, \\
13  &  $450 \div 500$    & 475 & 476.1 &\, 6.07 $\pm$ 2.05 \,\\
14  &  $500 \div 550$    & 525  & 526.0 &\, 8.49 $\pm$ 4.27 \, \\
15  & $550 \div 700$     & 625 & 632.9 &\, 17.4 $\pm$  10.3 \, \\
\hline
\end{tabular}
\label{tab:TFF}
\end{table}

The systematic uncertainty consists of two major contributions: the first due to the experimental resolution of the variables to which the analysis cuts are applied, 
and the second associated to the background fitting procedure. 

The systematic uncertainty due to the analysis cuts is evaluated moving by $\pm 1 \sigma$ all the variables on which a selection is applied. Cuts are moved 
once at a time, logging the deviation of counts in each bin of $\sqrt{q^2}$ from the original one. The relative deviations of counts coming from the different cuts are 
then summed bin by bin in quadrature to get the total relative uncertainty. When a variable is selected within a window, its edges are always moved oppositely in 
order to make the window wider or narrower according to the resolution. The resulting fractional uncertainty is of a few percent in most of the bins of lower $\sqrt{q^2}$, 
increasing up to 20\% in some of  the bins of higher 4-momentum transfer. There is no evidence of a single dominant cut with respect to the others; the contribution 
of the various analysis cuts is different for each bin of $\sqrt{q^2}$.

The systematic error associated to the fitting procedure is evaluated computing the deviation of the yield of the background function, with respect to the nominal one, 
when each of the four parameters is moved by $\pm 1 \sigma$ while fixing the other ones according to the correlation matrix. The four contributions thus obtained are summed 
in quadrature to get the total uncertainty on the background yield in each bin of $\sqrt{q^2}$. This error contribution is then propagated to $F_{\phi \pi^0} (q^2)$ through the 
number of signal candidates in each bin, which enters in the computation. The contribution in each bin of $\sqrt{q^2}$ is of a few percent.

\begin{figure}[!htb]
  \begin{center}
    \epsfig{file=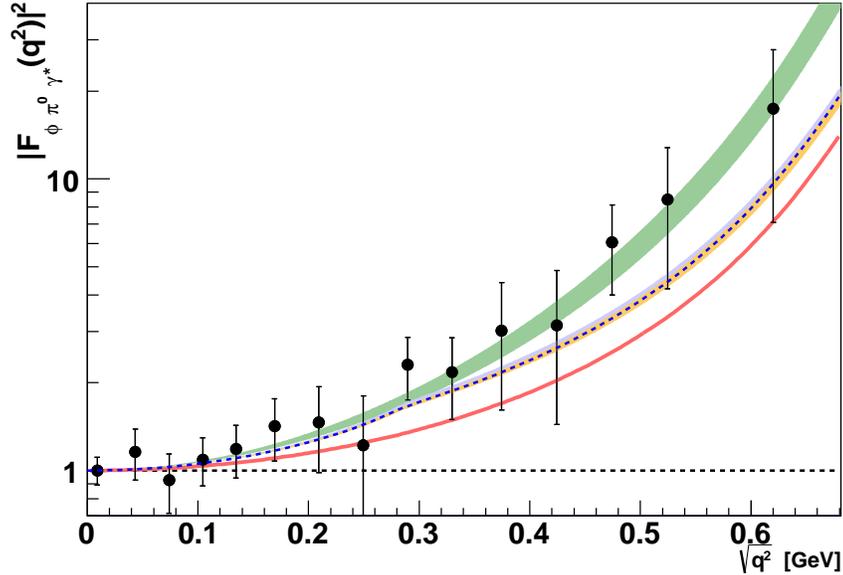,width=0.9\textwidth}
  \end{center}
  \caption{Comparison between the measurement of $| F_{\phi \, \pi^0}(q^2) |^2$ (black points) and the theoretical 
     predictions for this quantity based on:  the dispersive analysis of Ref. \cite{bib:SKN} (orange and cyan bands) and 
     Ref. \cite{bib:dan} (blue dashed line), the chiral theory approach of Ref. \cite{bib:iva} (green band), and the one-pole 
     VMD model (solid red line) (see Eqs. (49) and (50) of Ref. \cite{bib:dan}).}
  \label{fig:kloeTFF}
\end{figure}

In Fig. \ref{fig:kloeTFF}, our results on $| F_{\phi \, \pi^0} (q^2) |^2$ are compared with three different theoretical predictions. The best agreement is obtained with 
the Unconstrained Resonant Chiral Theory (UChT), with parameters extracted from a fit of the NA60 data \cite{bib:iva}.
We note that, as a consequence of the steepness and nonlinearity of the $e^+e^-$ invariant-mass spectrum, the TFF measured in a $\sqrt{q^2}$ bin cannot be 
associated to the corresponding bin center. For this reason, each experimental point of Fig. \ref{fig:kloeTFF} is associated with a $\sqrt{q^2}$ value weighted 
according to the theoretical shape predicted by UChT (see column labeled ``$\sqrt{q^2}$ UChT'' in Table \ref{tab:TFF}). As shown in Tab. \ref{tab:TFF}, with the 
given bin widths, the bin center is a good approximation of the weighted $\sqrt{q^2}$ in each bin, with the exception of the very first bin, where the $m_{e^+e^-}$ 
function is steeper. 

The transition form factors are often represented by a simple, VMD-inspired, one-pole parametrization:
\begin{equation}
\label{eq:1pole}
F (q^2) = \frac1{1 - q^2/\Lambda^2},
\end{equation}
from which the form factor slope parameter is obtained:
$$
b = \frac{\dd F (q^2)}{\dd q^2} \bigg|_{q^2 = 0} = \Lambda^{-2}.
$$
By fitting our data according to \eqref{eq:1pole}, we get $b_{\phi \pi^0} = (2.02 \pm 0.11)$ GeV$^{-2}$, to be compared with the one-pole approximation expectation, 
$b_{\phi \pi^0} = M_\phi^{-2}$, and the prediction of the dispersive analysis, $b_{\phi \pi^0} = (2.52 \cdots 2.68)$~GeV$^{-2}$, of Ref. \cite{bib:SKN}.

\subsection{Measurement of BR($\phi \to \pi^0 e^+e^-$)}
The branching ratio of the $\phi \to \pi^0 e^+e^-$ decay was obtained from the background-subtracted $e^+e^-$ mass spectrum by applying an efficiency correction 
evaluated bin by bin:
\begin{equation}
\text{BR}\,(\phi \to \pi^0e^+e^-) = \frac{\sum_i N_i / \epsilon_i}{\sigma_\phi \times \mathcal{L}_{\rm{int}} \times \text{BR}\,(\pi^0 \to \gamma\gamma)}, 
\label{eq:BRformula}
\end{equation} 
where $\sigma_\phi$ is the effective $\phi$ production cross-section, $\sigma_\phi = (3310\,\pm\,120)$ nb \cite{bib:eta}, 
$\mathcal{L}_{\rm{int}} = (1.69\,\pm\,0.01)$ fb$^{-1}$ \cite{bib:lum} is the integrated luminosity of data, and $\rm{BR}\,(\pi^0 \to \gamma\gamma)$ the branching 
ratio of $\pi^0$ into two photons \cite{bib:pdg}. $N_i$ is the number of signal candidates in the $i^{\,th}$ bin of $\sqrt{q^2}$ and $\epsilon_i$ is the corresponding 
selection efficiency, evaluated as the number of MC signal events in the $i^{\,th}$ bin after all the analysis steps, divided by the number of the corresponding 
generated events. The result covers the range $\sqrt{q^2} < 700$ MeV (the upper edge of the higher bin of $\sqrt{q^2}$) and is equal to:
\begin{equation}
\rm{BR}(\phi \to \pi^0 e^+e^-; \sqrt{q^2} < 700\;\rm{MeV}) = (1.19\,\pm\,0.05^{\,\,+0.05}_{\,\,-0.10}\,) \times 10^{-5}.
\label{eq:BR1}
\end{equation} 
Here, the first error results from the combination of the statistical one (2.2 \% in fraction) with the above quoted uncertainties on $\sigma_\phi$ and 
$\mathcal{L}_{\rm int}$. The second is a systematic one due to the analysis cuts and background subtraction (see sec. \ref{s:tff}). The error on $\epsilon_i$ due 
to the parametrization of the TFF in the MC is negligible.

The result can be extended to the full $\sqrt{q^2}$ range evaluating the fraction of the integral in the $e^+e^-$ invariant-mass spectrum which is not covered 
by the analysis. The extrapolation has been computed according to the theoretical model that best fits the data \cite{bib:iva}. The estimate of the total 
branching ratio is:
\begin{equation}
\rm{BR}\,(\phi \to \pi^0 e^+e^-) = (\,1.35  \pm  0.05^{\,\,+0.05}_{\,\,-0.10}\,) \times 10 ^{-5}.
\label{eq:BR2}
\end{equation} 
This result improves the previous measurements by SND and CMD-2 experiments and is in agreement with the theoretical predictions shown in 
Table \ref{tab:expt}.
\begin{table}[!ht]
\centering
\caption{Previous determination of BR ($\phi \to \pi^0 e^+e^-$) by SND \cite{bib:SND} and CMD-2 \cite{bib:CMD2}. 
The PDG average is $(1.12 \pm 0.28) \times 10^{-5}$ \cite{bib:pdg}. The theoretical predictions are also reported. For 
Ref. \cite{bib:SKN}  ``once'' (``twice'') refers to the dispersive analysis with one (two) subtractions.}
\bigskip
\begin{tabular}{ c c c }
\hline  
\hline
  & & BR $(\phi \to \pi^0 e^+e^-) \times 10^5$  \\ 
\hline
Experiment  & SND & $1.01 \pm 0.28 \pm 0.29$  \\
 & CMD-2 & $1.22 \pm 0.34 \pm 0.21$  \\
\hline
Theory & Schneider et al. \cite{bib:SKN} (``once'') & (1.39 \ldots 1.51) \\ 
 & Schneider et al. \cite{bib:SKN} (``twice'') & (1.40 \ldots 1.53) \\ 
 & Danilkin et al. \cite{bib:dan} & 1.45 \\
\hline
\hline
\vspace*{0.2cm}
\end{tabular}
\label{tab:expt}
\end{table}

\section{Conclusions}
Analyzing the conversion decay $\phi \to \pi^0 e^+e^-$, we measured for the first time the modulus square of the $F_{\phi \pi^0}$ transition 
form factor for $\sqrt{q^2}$ below 700~MeV. The data are in agreement with the theoretical prediction based on the Unconstrained Resonant 
Chiral Theory (UChT), with parameters extracted from a fit of the NA60 data. From the same data set we obtained a value of 
$\rm{BR}\,(\phi \to \pi^0 e^+e^-; \sqrt{q^2} < 700\;\rm{MeV}) = (1.19 \,\pm\,0.05^{\,\,+0.05}_{\,\,-0.10}\,) \times 10^{-5}$. An extrapolation based 
on the theoretical model in agreement with the data has been used to extend the result to the full $\sqrt{q^2}$ range. The value obtained is 
$\rm{BR}\,(\phi \to \pi^0 e^+e^-) = (1.35 \,\pm\,0.05^{\,\,+0.05}_{\,\,-0.10}\,) \times 10^{-5}$, that improves significantly the results obtained by 
SND and CMD-2 experiments, and is in  agreement with theoretical predictions. 

\section*{Acknowledgments}
We warmly thank our former KLOE colleagues for the access to the data collected during the KLOE data taking campaign.
We thank the DA$\Phi$NE team for their efforts in maintaining low background running conditions and their collaboration during 
all data taking. We want to thank our technical staff: G.F. Fortugno and F. Sborzacchi for their dedication in ensuring efficient 
operation of the KLOE computing facilities; M. Anelli for his continuous attention to the gas system and detector safety; 
A. Balla, M. Gatta, G. Corradi and G. Papalino for electronics maintenance; M. Santoni, G. Paoluzzi and R. Rosellini for general 
detector support; C. Piscitelli for his help during major maintenance periods. 
We thank Prof. B. Kubis and Dr. I. Danilkin for the detailed result of the calculation of Refs. \cite{bib:SKN} and \cite{bib:dan}, respectively. 
We are also very grateful to Dr. S. Ivashyn for providing us the formula for $\cos \theta^*$ and for the many enlightening discussions 
during all the phases of the analysis. 
This work was supported in part by the EU Integrated Infrastructure Initiative Hadron Physics Project under contract number 
RII3-CT- 2004-506078; by the European Commission under the 7th Framework Programme through the `Research Infrastructures' 
action of the `Capacities' Programme, Call: FP7-INFRASTRUCTURES-2008-1, Grant Agreement No. 227431; by the Polish National 
Science Centre through the Grants No.\
2011/03/N/ST2/02652,
2013/08/M/ST2/00323,
2013/11/B/ST2/04245,
2014/14/E/ST2/00262,
2014/12/S/ST2/00459.


\end{document}